\documentclass[aps, prx, reprint, longbibliography, nofootinbib,superscriptaddress, floatfix]{revtex4-2}
\pdfoutput=1

\usepackage{slashed}
\usepackage{verbatim}
\usepackage[T1]{fontenc}
\usepackage[utf8]{inputenc}
\usepackage[colorlinks]{hyperref}
\usepackage{mathbbol}
\usepackage[dvipsnames]{xcolor}
\usepackage[normalem]{ulem}
\usepackage{svg}
\usepackage[english]{babel}
\usepackage{lipsum}
\usepackage{physics}
\usepackage{dcolumn}
\usepackage{tensor}
\usepackage{comment}
\usepackage{graphicx,color,overpic,mathtools}
\usepackage{amsthm,amsmath,amssymb,hyperref,mathrsfs}
\usepackage{braket,bm,bbm,setspace}
\usepackage{cancel}
\usepackage{float}
\hypersetup{
    colorlinks=false,
    pdfborder={0 0 0},
}

\begin{document}

\title{Spin dependence of black hole ringdown nonlinearities}

\author{Jaime Redondo--Yuste}
\email[]{jaime.redondo.yuste@nbi.ku.dk}
\affiliation{Niels Bohr International Academy, Niels Bohr Institute, Blegdamsvej 17, 2100 Copenhagen, Denmark}
\author{Gregorio Carullo}
\affiliation{Niels Bohr International Academy, Niels Bohr Institute, Blegdamsvej 17, 2100 Copenhagen, Denmark}
\author{Justin L. Ripley}
\affiliation{Illinois Center for Advanced Studies of the Universe \& Department of Physics, University of Illinois at Urbana-Champaign, Urbana, Illinois 61801, USA}
\author{Emanuele Berti}
\affiliation{William H. Miller III Department of Physics and Astronomy, Johns Hopkins
University, 3400 North Charles Street, Baltimore, Maryland, 21218, USA}
\author{Vitor Cardoso}
\affiliation{Niels Bohr International Academy, Niels Bohr Institute, Blegdamsvej 17, 2100 Copenhagen, Denmark}
\affiliation{CENTRA, Departamento de F\'{\i}sica, Instituto Superior T\'ecnico -- IST, Universidade de Lisboa -- UL,
Avenida Rovisco Pais 1, 1049-001 Lisboa, Portugal}

\begin{abstract}
The nonlinear character of general relativity leaves its imprint in the coalescence of two black holes, from the inspiral to the final ringdown stage. To quantify the impact of nonlinearities, we work at second order in black hole perturbation theory and we study the  excitation of second-order modes relative to the first-order modes that drive them as we vary the black hole spin and the initial data for the perturbations. 
The relative amplitude of second-order modes is only mildly dependent on the initial data that we consider, but it strongly decreases for large black hole spins. 
This implies that the extrapolation of calculations based on the Kerr-CFT correspondence to subextremal Kerr black holes should be viewed with caution.
\end{abstract}

\maketitle

%
\noindent \textbf{\emph{Introduction.}} The study of strong gravitational fields may unlock some of the most persistent puzzles in our understanding of fundamental interactions. The detection of gravitational waves from the merger of compact objects~\cite{LIGOScientific:2016aoc,AdvLIGO, AdvVirgo,Kagra} turned strong gravity into an active and data-driven research field, allowing for new tests of general relativity (GR) in the nonlinear regime and, in particular, of black hole (BH) physics~\cite{Berti:2015itd,Barack:2018yly,LIGOScientific:2016lio,LIGOScientific:2019fpa,LIGOScientific:2020tif,LIGOScientific:2021sio}. 

An especially appealing dynamical setting concerns the late stages of the coalescence of compact binaries, when the final BH relaxes to a stationary Kerr state, ``ringing down'' in a set of characteristic quasinormal modes (QNM)~\cite{Kokkotas:1999bd,Berti:2009kk,Konoplya:2011qq}. The ``black hole spectroscopy'' program is concerned with the details of this process, and it requires precise theoretical predictions to be compared against data~\cite{Berti:2005ys, Berti:2016lat}. The frequencies of the QNMs are a benchmark prediction of GR, and they have been used to carry out consistency tests with LIGO-Virgo-KAGRA data~\cite{LIGOScientific:2016lio,Brito:2018rfr,Carullo:2019flw,LIGOScientific:2020tif,Carullo:2021dui,Carullo:2021oxn,LIGOScientific:2021sio,Cotesta:2022pci,Silva:2022srr,Finch:2022ynt,Crisostomi:2023tle}. 

Most prior work on ringdown was done within linear perturbation theory, which is valid when the backreaction of the perturbations on the spacetime is negligible. However, right after the merger of two BHs backreaction is significant, and nonlinear phenomena have been identified in several studies (see e.g.~\cite{Zlochower:2003yh, Abdalla:2006vb, London:2014cma,Sberna:2021eui,Cheung:2022rbm, Mitman:2022qdl, Baibhav:2023clw}). In order to achieve high-precision ringdown tests, it is critical to understand the transition between the nonlinear and the linear regime. 
One example of a nonlinear effect are quadratic quasinormal modes (QQNMs): loosely speaking, these are driven, quadratic (and therefore, nonlinear) combinations of QNMs. The presence of these driven modes is predicted by second-order perturbation theory~\cite{Gleiser:1995gx,Brizuela:2009qd,Ioka:2007ak,Nakano:2007cj,Pazos:2010xf,Loutrel:2020wbw,Ripley:2020xby,Lagos:2022otp}, and they play an important role in gravitational waveforms emitted by binary BH mergers~\cite{London:2014cma,Cheung:2022rbm, Mitman:2022qdl}.  
Including these effects may be crucial to model certain high angular ringdown modes~\cite{Yang:2014tla}.

In a spherical decomposition of the  gravitational waveform, QQNMs are generated from linear QNMs through the channel $(\ell m) \times (\ell^\prime m^\prime) \rightarrow (L  M)$, with standard angular selection rules such that $M = m + m'$.
Recently, the amplitude of the $L=M=4$ QQNM was measured by fitting binary BH merger simulations in numerical relativity, 
and compared with the squared amplitude of the \emph{parent} mode 
$\ell = \ell^\prime = m = m^\prime = 2$~\cite{Cheung:2022rbm, Mitman:2022qdl}. 
Focusing on the fundamental ($n=0$) QNM contribution, both references obtained a similar value for the ratio 
\begin{equation}\label{eq:ratio-reported}
    \mathcal{R}_{220\times220} = \frac{A_{220\times 220}}{A_{220}^2} \sim 0.15 - 0.2  \,  , 
\end{equation}
where $A_{\ell m n}$ denotes the amplitude of the mode with angular numbers $\ell m$ and overtone index $n$ in a spin-weighted spheroidal harmonics decomposition (``parent mode''), and $A_{220\times 220}$ denotes the amplitude of the QQNM resulting from the nonlinear interaction of a pair of $220$ modes (``child mode''). 

A similar ratio was recently borne out of calculations using the Kerr/CFT${}_2$ correspondence ~\cite{Kehagias:2023ctr} (see also Ref.~\cite{Guerreiro:2023gdy}), which assumes near-extremal BHs~\cite{Guica:2008mu}. 
This result raises an interesting question: \emph{is the ratio~\eqref{eq:ratio-reported} generically independent of the characteristics of the perturbed Kerr BH, including its spin and the initial conditions of the perturbations?} 
Here we show that it is not.

\noindent \textbf{\emph{Set-up.}}
\begin{figure}
    \centering
    \includegraphics[width = 0.9\columnwidth]{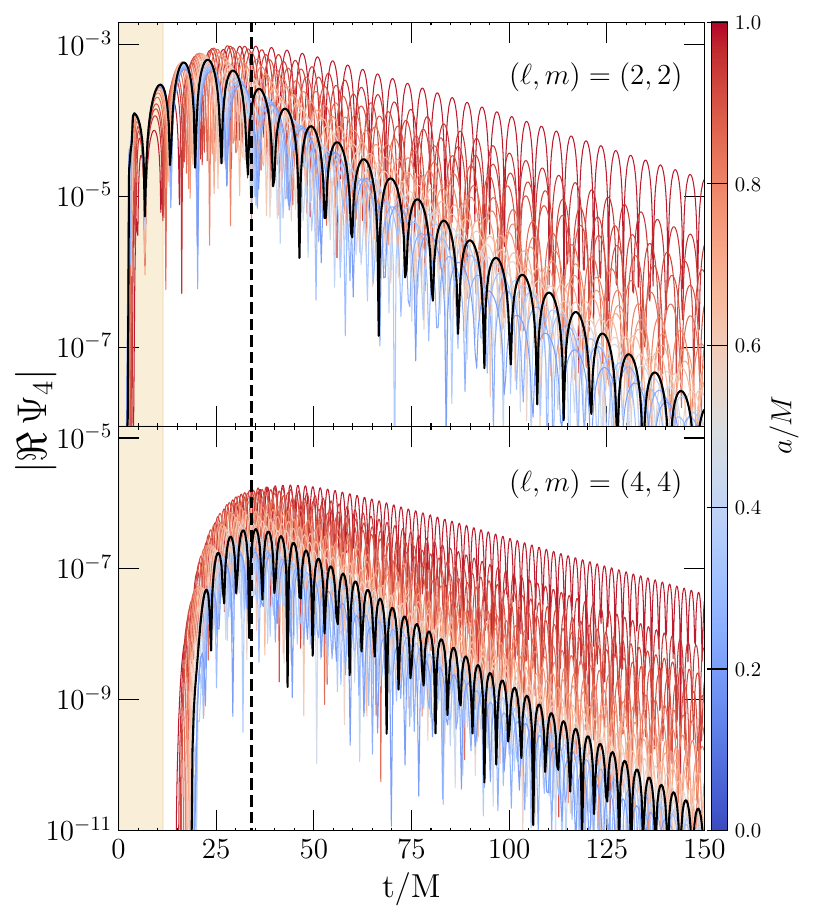}
    \caption{Absolute value of the real part of the perturbation scalar $\Psi_4^{(1/2)} = \mathcal{T}[h_{ab}^{(1/2)}]$ in the linear (top) and quadratic (bottom) sectors, for the angular modes indicated in the legend. 
    Lines correspond to perturbations of a Kerr background with different spin $a/M$ (different colors), excited by an initial profile with $\{ A, r_l/r_h, r_u/r_h \} = \{ 0.01, 1.5,  2.5 \}$, where $r_h \equiv M+\sqrt{M^2-a^2}$.
    For the specific spin value $a/M=0.5$ (black line) we show the reference timewhen the quadratic mode attains its maximum (vertical dashed line), and we also highlight in beige the region where constraint violations due to the initial data are present.
    }
    \label{fig:Psi4_Time_Spin}
\end{figure}
We study the evolution of a perturbed BH with mass $M$ and angular momentum $J=Ma$ up to second order , with line element 
\begin{equation}
g_{ab} = \bar{g}_{ab} + \varepsilon h^{(1)}_{ab} + \varepsilon^2 h^{(2)}_{ab}\, ,
\end{equation}
where $\bar{g}_{ab}$ is the background Kerr metric, $h^{(n)}_{ab}$ is the $n$--th order perturbation, and $\varepsilon$ a perturbative parameter. The vacuum Einstein equations reduce to one master Teukolsky equation at each perturbative order~\cite{Teukolsky:1973ha,Campanelli:1998jv,Loutrel:2020wbw}. Up to second order, the system of equations is 
\begin{equation}
\mathcal{O}_{-2} \mathcal{T} [h^{(1)}_{ab}] =  0\,,\quad
        \mathcal{O}_{-2} \mathcal{T} [h^{(2)}_{ab}] = \mathcal{S}[h^{(1)}_{cd}]    \,  ,\label{eq:master-eqs}
\end{equation}
where $\mathcal{O}_{-2}$ is the Teukolsky operator, and $\mathcal{T}$ is an operator that relates the metric perturbation to the Weyl scalar $\Psi_4$. The second-order equation is sourced by first-order fluctuations $\mathcal{S}[h^{(1)}_{cd}]$.
We solve the Teukolsky equation $\mathcal{O}_{-2}\Psi_4^{(1)}=0$ to obtain $\Psi_4^{(1)} = \mathcal{T}[h_{ab}^{(1)}]$, which at future null infinity can be directly related to the gravitational-wave strain by 
\begin{equation}\label{eq:psi_h}
    \Psi_4 \xrightarrow{\mathscr{I}^+} -2\Bigl(\ddot{h}_+ - i \ddot{h}_\times\Bigr)    \,  ,
\end{equation}
where $h_{+, \times}$ are the two gravitational-wave polarizations (the prefactor of $2$ is sensitive to the code's tetrad conventions~\cite{Ripley:2020xby,teuk-fortran-2020,Hengrui}).
Knowledge of the full $h_{ab}^{(1)}$ is not necessary to study gravitational radiation at first order, but it is required to build the source term to compute second-order quantities. The procedure to obtain $h_{ab}^{(1)}$ from $\Psi_4^{(1)}$ is known as metric reconstruction, and it is a topic of active research~\cite{Wald:1978mtr, Campanelli:1998jv, Loutrel:2020wbw,Green:2019nam, Spiers:2023cip}.
We reconstruct the metric $h_{ab}^{(1)}$ from $\Psi_4^{(1)}$ by directly integrating the Einstein equations in the Newman-Penrose formalism~\cite{Loutrel:2020wbw}.

We numerically solve the system of equations~\eqref{eq:master-eqs} with the code described in Refs.~\cite{Ripley:2020xby,teuk-fortran-2020}. The code solves the Teukolsky equation and metric reconstruction equations on a horizon-penetrating, hyperboloidally compactified domain -- i.e., it 
using coordinates stretching from the BH horizon to future null infinity on constant-time hypersurfaces. Initial data for $\Psi_4^{(1)}$ is given by:
\begin{eqnarray}
\label{eq:initial-data-psi4}
\Psi_4^{(1)}(t=0) &=& \phi_0{}\,\,_{-2} P_\ell^m(\vartheta) e^{im\phi}    \,  ,   \\
\phi_0&=& A \Bigl(\frac{r-r_l}{r_u-r_l}\Bigr)^2\Bigl(\frac{r_u-r}{r_u-r_l}\Bigr)^2\nonumber \\
&\times& \mathrm{exp}\Bigl(-\frac{1}{r-r_l}-\frac{2}{r_u-r}\Bigr) \theta(r-r_l)\theta(r_u-r)    \,  .\nonumber
\end{eqnarray}
Above, ${}_sP_\ell^m$ denotes the spin-weighted associated Legendre function, the numbers $\ell$ and $m$ are constants which characterize the angular structure of the initial data, and $\theta$ denotes the Heaviside step function.
The parameters $\{A,r_l,r_u\}$ characterize the initial profile, fixing respectively its amplitude, lower and upper radius. 
The initial velocity profile, $\partial_t\Psi_4^{(1)}$, is chosen so that the pulse moves towards the BH. 
We additionally set $h_{ab}^{(1)}\left(t=0\right)=0$. While these initial data do not satisfy the Einstein constraint equations, so long as $|m|\geq2$ the constraint violating modes propagate off of our computational domain in  a finite amount of time $T_R$~\cite{Ripley:2020xby}. We set $\Psi_4^{(2)}=0$, $h_{ab}^{(2)}=0$, which satisfies the second-order Einstein constraint equations.
Numerical convergence is discussed in the Supplemental Material (SM).
As the numerical grid represents a hyperboloidal slicing of the Kerr background, we can extract $\Psi_4^{(1)}$ and $\Psi_4^{(2)}$ at null infinity.
We then project those quantities onto spin-weighted spherical harmonics. 

In Fig.~\ref{fig:Psi4_Time_Spin} we show how the perturbations extracted at null infinity depend on the BH spin.
The BH is initially perturbed by an incoming $\ell = m = 2$ mode (upper panel). 
The lower panel shows the $\ell = m = 4$ mode, excited due to the nonlinear contributions of the initial perturbation. 
Constraint violations have no significant impact on our ability to study second-order modes.

\noindent \textbf{\emph{Ringdown model.}} 
To compare our results with Refs.~\cite{Kehagias:2023ctr,Cheung:2022rbm, Mitman:2022qdl}, we study the strain amplitude $h$ instead of $\Psi_4$.
Our numerical simulations end before we observe a power-law tail, and the modeling starting time imposed by the metric reconstruction implies that the initial burst should not affect our calculations.
Hence, for each of the spherical modes we consider a template consisting of a superposition of damped sinusoids
\begin{eqnarray}
    h_{\ell  m} (t) &=& \sum_{\ell^\prime=2}^{\infty} \sum_{n=0}^{\infty} \left[ \mathcal{A}_{\ell^\prime m n} \, e^{i(\omega_{\ell^\prime m n}(t-t_{0})+\phi_{\ell^\prime m n})} \right. \nonumber\\
    &+& \left. \mathcal{A}_{\ell^\prime -m n} \, e^{i(\omega_{\ell^\prime -m n}(t-t_{0})+\phi_{\ell^\prime -m n})} \right] \, ,\label{eq:Kerr_model}
\end{eqnarray}
with $t> t_0$ and the complex frequencies $\omega_{\ell^\prime m n}$ correspond to the QNMs of a perturbed Kerr BH.\footnote{
The frequencies correspond to a Kerr BH with mass and spin fixed by the initial conditions. Changes in the mass and spin are proportional to $(h^{(1)}_{ab})^2$, and thus, they do not affect the evolution of $\Psi^{(1)}$ and $\Psi^{(2)}$ to second order in perturbation theory. In general, higher-order perturbations (starting at cubic order, i.e., $\Psi^{(3)}$) would be sensitive to this effect~\cite{Sberna:2021eui}.
}
All amplitudes and phases will be defined with respect to the arbitrary reference time $t_0=0$, which we choose to coincide with the peak of the second-order perturbation (see Fig.~\ref{fig:Psi4_Time_Spin}).
The sum on $\ell^\prime$ is due to spherical-spheroidal mode mixing~\cite{Berti:2005gp,Buonanno:2006ui,Kelly:2012nd,Berti:2014fga,London:2018nxs}. We compute the mixing coefficients using the Black Hole Perturbation Toolkit~\cite{BHPToolkit}, as described in the SM.
The first term corresponds to ``corotating'' (prograde) modes with $\Re{\omega_{\ell  m  n}} > 0$, while the second corresponds to ``counterrotating'' (retrograde) modes with $\Re{\omega_{\ell  -m n}} < 0$, which are increasingly suppressed for large values of $a/M$~\cite{Berti:2005ys,Lim:2019xrb,Li:2021wgz}.
As our model assumes a QNM-driven regime, we can analyze $\psi_{4, \ell m}$ and reconstruct the strain amplitudes by a simple frequency rescaling: $\mathcal{A}_{\ell m n} = (2\omega_{\ell  m  n}^2)^{-1} \mathcal{A}^{\Psi}_{\ell  m  n}$, as dictated by Eq.~\eqref{eq:psi_h} at future null infinity. This rescaling is in good agreement with an alternative calculation in which we compute the strain $h$ through two direct integrations in time of $\Psi_4$, and then extract the amplitude of $h$, but it avoids the integrations (which increase numerical noise).

We refer to each of the two contributions to a given $(\ell m)$ multipole of $h$ as a ``mode,'' each with an amplitude $\mathcal{A}_{\ell m n}$ and a phase $\phi_{\ell m n}$, treated as two unknown parameters to be inferred from the numerical data.
%

\noindent \textbf{\emph{Bayesian mode extraction.}} 
We explored a wide range of initial conditions and spins.
Properly accounting for errors on the fitting parameters is crucial, because some of the subdominant contributions can be comparable to the numerical resolution (this is true, e.g., for counterrotating modes and near-extremal spins).
For this reason, we address the fitting problem through Bayesian inference~\cite{Jaynes2003}.
In this way the numerical error is folded in the inferred parameters, yielding not only a point estimate of the quantities of interest, but the full information available at a given resolution (including the uncertainties).

The complex QNM frequencies are uniquely determined by the Kerr BH mass (which enters only as a scaling factor) and spin of each simulation.
We choose uniform priors on all sampling parameters within the ranges $\ln \mathcal{A} \in [-15, 0]$ for the linear contributions, $\ln \mathcal{A} \in [-35, -5]$ for the quadratic contributions, and $\phi \in [0, 2\pi]$. 
The dominant uncertainty in the reconstruction is introduced by the resolution error, so we consider a gaussian likelihood, with an error at each time given by the difference between the two highest resolutions available.
This choice guarantees a conservative and unbiased mode extraction, as we verified by simulating and recovering mock ringdown signals superimposed to the numerical error.
To perform the extraction, we have developed a dedicated \texttt{python} package, \texttt{bayRing}, which we make publicly available~\cite{public_release}.
The waveform model is interfaced through \texttt{pyRing}~\cite{pyRing,Carullo:2019flw}, the QNM frequencies are computed through the \texttt{qnm}~\cite{Stein:2019mop} package, and the high-dimensional parameter space exploration is achieved through the parallel nested sampling algorithm~\cite{skilling2006}, as implemented in \texttt{raynest}~\cite{raynest}.
We use 128 live points, and 256 maximum Markov Chain Monte Carlo (MCMC) steps. 
For each analysis, we run 4 parallel explorations, and combine them in a unique posterior distribution by weighting them through their Bayesian evidence. 
This last step is crucial to ensure that our estimate does not depend on the random initial seed (the most challenging fitting models contain up to $12$ free parameters).

In the SM we discuss how we take into account the starting time uncertainty (see~\cite{Dorband:2006gg, London:2018gaq}) and possible additional systematic model uncertainties in our error budget,
as well as the procedure used to select a ``default set'' of modes in our analysis. These modes are the $220, 2\text{-}20,320, 221$ for the linear order, and the $440, 4\text{-}40$ and $220\times 220$ for the second order perturbations. 

Roughly speaking, our model and inference strategy are considered successful if the analysis residuals are smaller than the error induced by the numerics, and if we observe a saturation in the Bayesian evidence when including additional modes.

\begin{figure}
    \centering
    \includegraphics[width = 0.9\columnwidth]{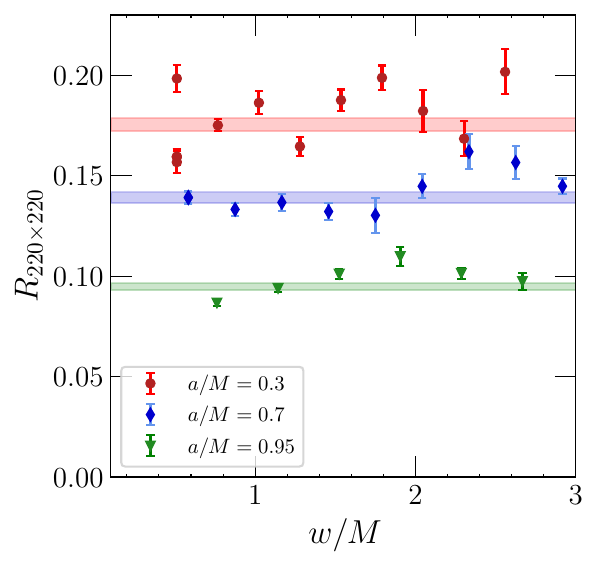}
    \caption{
    Ratio $\mathcal{R}_{220\times220} = A_{220\times220}/A^2_{220}$ as a function of the width of the initial pulse $w = r_u - r_l$, for a BH spin $a/M = [0.3,0.7,0.95]$ (in red, blue, green, respectively) and $r_l = 1.5r_h$. Error bars indicate uncertainties due to numerical noise, the fitting procedure, mode selection, and the time shift problem. 
    The average value (with its associated standard deviation) is shown as a colored shaded region. When a given value of $w$ represents multiple points, those are computed using different values of $r_l$, further showing that the ratio depends only very mildly on the initial conditions. 
    }
    \label{fig:ICs}
\end{figure}

\begin{figure}
    \centering
    \includegraphics[width =  \columnwidth]{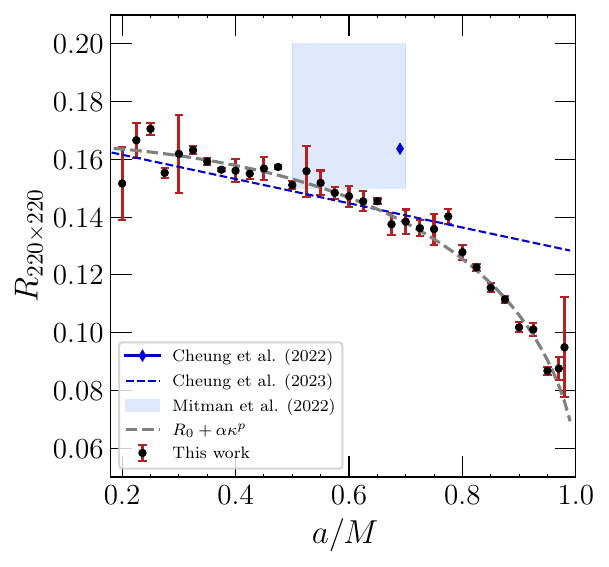}
    \caption{Ratio $\mathcal{R}_{220\times220}$, as a function of the spin of the remnant BH. The value decreases with the spin, especially as one approaches the extremal limit. 
    The gray dashed line represents the best fit of the data points assuming a power law in terms of the surface gravity $\kappa$, $\mathcal{R}_{220\times 220} = R_0 + (M\kappa)^p$. The best fit parameters are given in Eq.~\eqref{eq:bestFit}.
    We also represent the values for the non--linear ratio $\mathcal{R}_{220\times220}$ obtained originally from NR in~\cite{Cheung:2022rbm} (referred to as Cheung et al.) and~\cite{Mitman:2022qdl} (referred to as Mitman et al.). The blue dashed line corresponds to the linear hyperfit reported in~\cite{Cheung:2023vki}, based on a catalogue of NR simulations. }
    \label{fig:Spins}
\end{figure}

\noindent\textbf{\emph{Dependence on the initial data.}} 
We have produced a catalog of runs with varying initial conditions and BH spins.
For simplicity, we always consider initial conditions in the spherical mode $(\ell,\, m ) = (2,\, 2)$, so that the dominant quadratic mode is the $2 2\times 2  2$. Our main results are summarized in Figs.~\ref{fig:ICs} and \ref{fig:Spins}.

An important question concerns the dependence of the nonlinear ratio $\mathcal{R}_{220\times220}$ on the initial data of Eq.~\eqref{eq:initial-data-psi4}. The overall amplitude $A$ of the initial perturbation is a scaling factor, hence it must factor out of the ratio, but there could be nontrivial dependence of this ratio on other parameters, such as the location of the inner boundary, $r_l$, and the ``width'' of the initial pulse, $w = r_u - r_l$. In Fig.~\ref{fig:ICs} we show that the ratio is only mildly dependent (if at all) on $r_l$ and $w$. The figure shows that the mild dependence on initial conditions is valid for all values of the spins. Notice that for smaller spins, due to the increased relevance of counterrotating modes, the uncertainties in the extraction are larger. 

The uncertainties in the extraction also become larger as the pulse becomes wider ($w \geq 3r_h$) or it is placed further away from the horizon ($r_l \geq 2r_h$). This is due to two reasons: the numerical evolution code uses hyperboloidal, compactified coordinates, and thus it needs more resolution to resolve features that are further away from the horizon or from null infinity. Moreover, initial pulses located far away correspond to a larger value of $T_R$, and this affects the time window of the signal available for the fit.
The range of initial conditions we explored is not large enough to conclude that $\mathcal{R}_{220\times220}$ is constant across the whole parameter space. 

The results reported in Fig.~\ref{fig:ICs} can be recovered analytically, assuming that the system is in a QNM-driven regime. Take the first-order perturbation to be a single mode: $\Psi_4^{(1)} = A_{220} f(R) e^{i\omega_{220}(t-t_0)}$, with $f(R)$ a normalized eigenfunction ($f(R=0)=1$) in terms of the compactified coordinate where $R=0$ corresponds to $\mathscr{I}^+$.
Then, the initial conditions only affect the value of $A_{220}$~\cite{Krivan:1997hc, Berti:2006wq}. The inhomogeneous part of the second-order perturbation is given by 
%
$    \Psi_4^{(2)}(t,R) = \int dR' G(R, R') \mathcal{S}[\Psi_4^{(1)}] \, ,$
%
where $G(R,R')$ is the radial Green's function associated to $\mathcal{O}_{-2}$, and $\mathcal{S}$ is the quadratic source term. 
Assuming a similar QNM structure for the quadratic perturbation (i.e., $\Psi_4^{(2)} = A_{220\times 220} g(R) e^{2i\omega_{220}(t-t_0)}$, with $g(R=0)=1$) we have 
\begin{equation}\label{eq:Ratio-theory}
    \mathcal{R}_{220\times 220} = \frac{A_{220\times 220}}{A_{220}^2} = \int dR' G(0,R') \mathcal{S}[f(R')] \, ,
\end{equation}
where the quadratic source term depends only on the radial profile of the linear mode. This suggests that the ratio $\mathcal{R}_{220\times 220}$ has no dependence on the initial conditions, as long as the nonlinear mode is generated exclusively via quadratic effects on a single mode. 
However, nonlocal behaviour of the source term may result in some nontrivial dependence on the initial conditions.

\noindent\textbf{\emph{Dependence on BH spin.}} 
The dependence on the spin can be studied in a similar way. We now assume the pulse to be at $r_l = 1.5r_h$, with width $w = r_h$.
The Green's function and the source term depend explicitly on the BH spin, thus Eq.~\eqref{eq:Ratio-theory} suggests that the nonlinear ratio will have a nontrivial spin dependence.

Our results are summarized in Fig.~\ref{fig:Spins}: there is no universality, and the ratio $\mathcal{R}_{220\times220}$ is indeed spin-dependent and decreases with spin, ranging roughly between $\mathcal{R}_{220\times220} =0.16$ for $a/M=0.2$ and $\mathcal{R}_{220\times220}=0.08$ for $a/M=0.98$. These variations are much larger than the estimated uncertainties, or than possible variations due to the mild dependence on the initial conditions. The values that we extract, moreover, are surprisingly close to the resulting nonlinear ratio reported in~\cite{Cheung:2022rbm, Mitman:2022qdl}. 
The spin-dependence of $\mathcal{R}_{220\times 220}$ is well fitted by a power law of the surface gravity $\kappa = \frac{\sqrt{M^2-a^2}}{2M (M + \sqrt{M^2-a^2})}$: our results are well fitted to an expression of the form $\mathcal{R}_{220\times 220}=R_0 + (M\kappa)^p$, with 
\begin{equation}\label{eq:bestFit}
    R_0=0.058  \pm 0.003 , \quad p = 1.61 \pm 0.03 \, .
\end{equation}
Uncertainties in the ratio extraction become smaller for rapidly spinning BHs, since at low spins counterrotating modes are as important as corotating ones. As a consequence, especially for the quadratic modes, our template should be enhanced with all the counterrotating contributions~\cite{Berti:2005ys,Lim:2019xrb,Li:2021wgz,Cheung:2023vki}, including the highly damped overtone and mode-mixing ones~\cite{Berti:2005gp,Buonanno:2006ui,Kelly:2012nd,Berti:2014fga,London:2018nxs}, severely complicating the inference problem and increasing computational cost. These contributions include, for example, mode-mixing terms such as the $640$ mode or the $541$ overtone, as well as additional quadratic terms such as $220\times 2\text{-}20$ mode. This is not of practical concern, since if universality does not happen already for this range of spins, there is no reason to expect universality to appear at lower spins. For this reason we do not include spins $a < 0.2$. 

\begin{figure}[t]
    \includegraphics[width = 0.8\columnwidth]{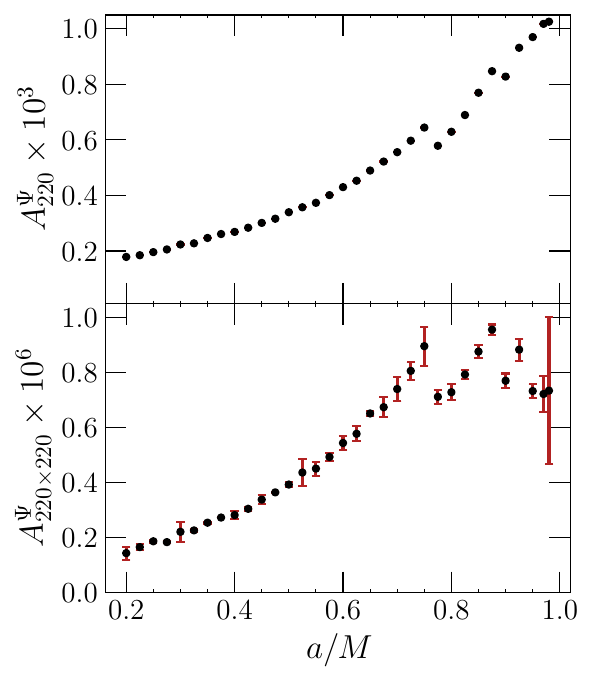}
    \caption{Amplitude of the fundamental $220$ mode (top) and of the quadratic $220\times 220$ mode (bottom) as a function of the BH spin. Notice that the amplitudes reported here refer to the master variable $\Psi_4$ and not to the GW strain.}
    \label{fig:Amplitudes}
\end{figure}

It is perhaps surprising that the nonlinear ratio $\mathcal{R}_{220\times 220}$ decreases as the BH spin increases, but it does not imply that nonlinearities are becoming smaller. In Fig.~\ref{fig:Amplitudes} we show that both the amplitude $\mathcal{A}_{220}$ of the linear and $\mathcal{A}_{220\times 220}$ of the nonlinear mode {\it grow} with BH spin, resulting in a ratio \eqref{eq:ratio-reported} which behaves as in Fig.~\ref{fig:Spins}. 
Since highly spinning BHs dissipate energy less efficiently, they could be more prone to nonlinearities.  The ratio that we study quantifies the strength of a process which transfers energy from low- to higher energy scales. If nonlinear turbulent dynamics suppresses it, in a manner similar to fluids in two spatial dimensions~\cite{Carrasco:2012nf, Green:2013zba, Adams:2013vsa, Yang:2014tla, Redondo-Yuste:2022czg}, it could explain a lower $\mathcal{R}_{220\times 220}$ at high spins.

Another feature apparent in Fig.~\ref{fig:Amplitudes} is a ``kink'' around $a/M\sim 0.75$. One possible explanation for this feature is the fact that on the equatorial plane ($\theta = \pi/2$), the light ring (which dominates QNM excitation) is inside the ergoregion exactly at $a/M = 1/\sqrt{2}\sim 0.7$. This could result in a qualitatively different behavior of the QNM excitation coefficients. Further clarification of this feature is left for future work.
%

\noindent \textbf{\emph{Conclusions.}} 
The full nonlinear content of Einstein's equations is currently accessible only via simulations in numerical relativity. However, essential insight can be obtained with various degrees of approximation. 
The next step in a BH spectroscopy program requires the understanding of second-order effects. We uncovered essential aspects of such higher-order terms, including a mild dependence on initial conditions, and a strong dependence on the BH spin. 


Remarkably, our results concerning the spin-dependence of the quadratic vs linear amplitudes ratio are in good agreement with the values reported from full nonlinear numerical simulations. 
Another recent study~\cite{Zhu:2024rej} explores the initial data and spin-dependence of the quadratic coefficient using both the second-order code employed here, and numerical relativity, albeit with a different mode inclusion criterion.
Close agreement is observed between the two studies, together with similar consistency with nonlinear simulations (c.f. our Fig~\ref{fig:Spins} with Fig.1 in~\cite{Zhu:2024rej}). Our results also show good agreement with semi--analytical calculations~\cite{Ma:2024qcv}.

Finally, the amplitude of first-order terms observed in our detectors is $h^{(1)}\sim 10^{-21}$, and one might be tempted to infer that $h^{(2)}\sim (h^{(1)})^2\sim 10^{-42}$. The correct scaling is instead $rh^{(2)}\sim (r h^{(1)})^2$ (one can see this directly in a gauge-invariant formalism, where second-order master wavefunctions $\Psi \propto rh$ are sourced by quadratic first-order terms~\cite{Gleiser:1995gx,Brizuela:2009qd,Ioka:2007ak,Nakano:2007cj}). Quadratic terms {\it are} important.

\noindent {\bf \em Acknowledgments.}
We thank Mark Ho-Yeuk Cheung, Alex Kehagias, Davide Perrone, and Toni Riotto for fruitful discussions. 
We are indebted to Hengrui Zhu for very useful discussions, and help in uncovering a data reading mistake in the initial version of this work.
G.C. thanks William East, Luis Lehner, Taillte May and Huan Yang for stimulating discussions, and Perimeter Institute for the generous hospitality during the last stages of this work.
J.L.R. thanks Hengrui Zhu for a helpful discussion on the tetrad normalization convention used in \texttt{SpEC}, and helpful conversations on nonlinear ringdown more generally.
J.R-Y, G.C. and V.C. acknowledge support by VILLUM Foundation (grant no. VIL37766) and the DNRF Chair program (grant no. DNRF162) by the Danish National Research Foundation.
G.C. also acknowledges funding from the European Union’s Horizon 2020 research and innovation program under the Marie Sklodowska-Curie grant agreement No. 847523 ‘INTERACTIONS’.
J.L.R. acknowledges support from the Simons Foundation through Award number 896696.
E.B. is supported by NSF Grants No. AST-2006538, PHY-2207502, PHY-090003 and PHY-20043, by NASA Grants No. 20-LPS20-0011 and 21-ATP21-0010, by the John Templeton Foundation Grant 62840, by the Simons Foundation, by the Amaldi Research Center, and by the Italian Ministry of Foreign Affairs and International Cooperation Grant No.~PGR01167.
E.B. also acknowledges the support of the Indo-US Science and Technology Forum through the Indo-US Centre for Gravitational Physics and Astronomy, grant IUSSTF/JC-142/2019.
V.C.\ is a Villum Investigator and a DNRF Chair.  
V.C. acknowledges financial support provided under the European Union’s H2020 ERC Advanced Grant “Black holes: gravitational engines of discovery” grant agreement no. Gravitas–101052587. 
Views and opinions expressed are however those of the author only and do not necessarily reflect those of the European Union or the European Research Council. Neither the European Union nor the granting authority can be held responsible for them.
This project has received funding from the European Union's Horizon 2020 research and innovation programme under the Marie Sklodowska-Curie grant agreement No 101007855.
This research was supported in part by Perimeter Institute for Theoretical Physics. Research at Perimeter Institute is supported in part by the Government of Canada through the Department of Innovation, Science and Economic Development Canada and by the Province of Ontario through the Ministry of Colleges and Universities.

\bibliography{biblio}

\clearpage


\renewcommand{\thesubsection}{{S.\arabic{subsection}}}
\setcounter{section}{0}

\section*{Supplemental material}
Here we provide additional information on the implementation of our Bayesian inference algorithm, \texttt{bayRing},
including spherical-spheroidal mode mixing (Sec.~\ref{app:mode_mixing}), the motivation for the ``default'' set of modes chosen for the inference in the main text (Sec.~\ref{app:default_modes}), our uncertainty estimates (Sec.~\ref{app:error_budget}), and the convergence of the numerical code (Sec.~\ref{app:code_convergence}). 

\subsection{Mode mixing}
\label{app:mode_mixing}
The Newman-Penrose scalar $\Psi_4$ extracted from numerical simulations is projected onto a spin-weighted spherical harmonic basis. However, the Teukolsky equation is only separable in spin-weighted spheroidal harmonics. Therefore, the natural basis used in the numerical implementation is different from the natural basis used for the extraction of the mode amplitudes. In order to overcome this difficulty, spherical-spheroidal mixing must be taken into account~\cite{Berti:2005gp,Buonanno:2006ui,Kelly:2012nd,Berti:2014fga,London:2018nxs}. We define the mixing coefficients $\alpha_{\ell'\ell m n}$ as the projection of a spheroidal harmonic $S_{\ell m}(\omega)$ with spheroidicity $\gamma = i a \omega$ and angular indices $(\ell,\, m)$ onto the spherical harmonic basis:
\begin{equation}
    S_{\ell m}(\omega_{\ell m n})
     = \sum_{\ell'=\ell}^\infty \alpha_{\ell' \ell m n} \, \, {}_{-2}Y_{\ell' m} \, .
\end{equation}
We use the Black Hole Perturbation Toolkit~\cite{BHPToolkit} implementation of the spin-weighted spheroidal harmonics to calculate the mixing coefficients. 

The presence of QQNMs produces a new kind of mixing coefficient, where the associated frequency is not given by a particular set of indices $(\ell,\, m,\, n)$, but it is rather the frequency associated to the QQNM. These mixing coefficients are associated to the expansion
\begin{equation}
    S_{LM}(\omega_{\ell_1 m_1 n_1} + \omega_{\ell_2 m_2 n_2}) = \sum_{L'}\beta^{L'M}_{\ell_1 m_1 n_1\times \ell_2 m_2 n_2} \, \, {}_{-2}Y_{L'M} \, ,
\end{equation}
where the usual selection rules apply. The mixing coefficients can be computed using the orthonormality of the spin-weighted spherical harmonics. They enter in the ratio of linear and quadratic amplitudes as follows: if $h_{\ell m n}$ denotes the amplitude extracted using Bayesian inference, and $\mathcal{A}_{\ell m n}$ denotes the actual amplitude of the expansion, then the physical ratio of interest is
\begin{equation}
    \mathcal{R}_{220\times 220} = \frac{\mathcal{A}_{220\times 220}}{\mathcal{A}_{220}^2} = \frac{h_{220\times 220}}{h_{220}^2} \frac{\alpha_{2220}^2}{\beta^{44}_{220\times 220}} \, .
\end{equation}
We take this factor into account whenever we extract the nonlinear ratio $\mathcal{R}_{220\times 220}$. 

\subsection{Construction of a default set of modes}
\label{app:default_modes}
The vast range of initial conditions and BH spins that we explore implies that some modes will be well above numerical uncertainties, while others (in particular, counterrotating modes) will be comparable to numerical uncertainties.
Including all such contributions in every case is computationally prohibitive, so we adopt a more practical approach.
Rather than assuming that a given set of modes is present in the data, we first agnostically extract the spectral content of the simulations.

We start by focusing on three representative simulations, with low-mid-high spins $a/M = \{0.3, 0.5, 0.7\}$, and analyze each of them (for the linear and quadratic sectors separately) using a ``free modes'' model:

\begin{equation}\label{eq:DS_model}
    \begin{aligned}
    h (t) = \sum_{i=0}^{N} \, \mathcal{A}_i \, e^{i[\omega_i(t-t_0) + \phi_i]} \, ,
    \end{aligned}
\end{equation}
where $\omega_{\mathrm{R}}^i = \mathrm{Re}(\omega_i)$, $\tau_i = 1/\mathrm{Im}(\omega_i)$, $A_i$, and $\phi_i$ are now all free parameters.
To ensure that we are in the QNM-driven regime, we set $t_0=10M$.
Note that $t_0=0$ corresponds to the peak of the quadratic mode, which is typically  $\sim 20M$ after the peak of the linear modes, ensuring that we are well into the QNM-driven regime.
Besides the standard settings stated in the main text, we choose additional priors on the frequency and damping time in the ranges $M\omega_{\mathrm{R}} \in [-2, 2]$, $\tau \in [1, 50]$.
To prevent ``mode switching'' during sampling, we order the modes by decreasing frequency.

As expected, the free modes posteriors latch onto the predicted Kerr frequencies (uniquely determined by the known mass and spin for each simulation) of several dominant contributions, e.g. $\{220, 2\text{-}20, 320, 3\text{-}20, 221 \}$ in the linear case.
All the modes we consider have a clear overlap with a single GR prediction, making their identification unambiguous.
We keep adding modes until the Bayesian evidence favors their inclusion. 
Equivalently, we stop adding modes once the residuals that we obtain are smaller than the error induced by the numerical resolution.
By this procedure we conclude that we must include $N=4$ modes.

Now that we have agnostically singled out the dominant contributions, we repeat the fit using Eq.~\eqref{eq:Kerr_model}.
However, now we fix the complex frequency to the predicted Kerr value and we vary the starting time $t_0 \in [0, 25]M$, in steps of $M$.
We start by fixing the dominant mode (the $220$ in the linear case) and we add the modes selected above one by one in order of decreasing amplitude, as determined by the free-frequency modes fit.
This procedure has proven to be successful in a wide range of QNM extraction problems, ranging from perturbation theory to full binary simulations~\cite{Baibhav:2023clw}.

\begin{figure}[t]
    \includegraphics[width = 0.9\columnwidth]{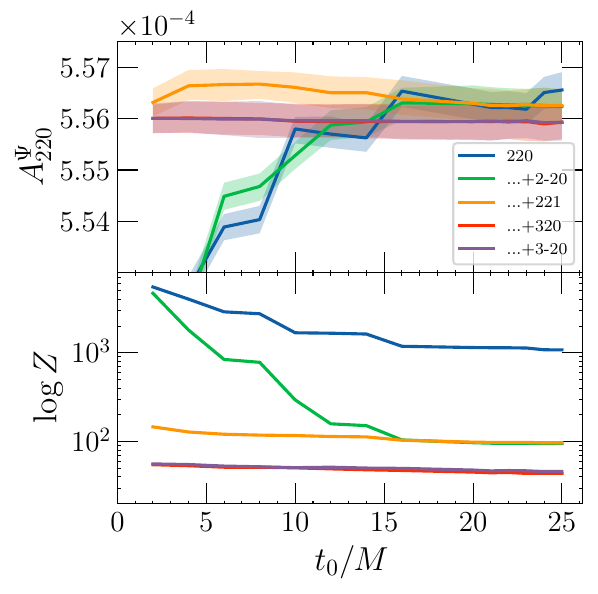}
    \caption{Amplitude of the $220$ mode (top) and Bayesian evidence $-\log Z$ as a function of the starting time of the fit, $t_0$. The BH spin is $a/M = 0.7$. Each line corresponds to a different mode content. The legend indicates the last mode added to the template (in order: $220, 2\text{-}20, 221, 320, 3\text{-}20$) and shaded regions represent the $90\%$ Bayesian inference confidence interval, which also accounts for numerical uncertainties. In the top panel the last two lines overlap, meaning that including one additional mode does not further change our extraction of the amplitude. The bottom panel additionally shows that including the $3\text{-2}0$ does not increase the evidence.}
    \label{fig:Agnostic_lin}
\end{figure}

In the top panel of Fig.~\ref{fig:Agnostic_lin} we show how the amplitude of the $220$ mode is impacted by the addition of other modes (as indicated in the legend).
When fitting using only the $220$ mode, the amplitude displays large variations at early times (shaded regions indicate the uncertainty induced by the numerical resolution).
When including the first overtone ($221$) mode, the amplitude variation is greatly reduced, and the $220$ late-time amplitude varies by $1\%$.
Adding the $320$ mode further reduces the amplitude variation. On the contrary, adding the next dominant mode $3\text{-}20$ has a negligible impact on the $220$ mode amplitude (which remains constant), and instead decreases the evidence very slightly (due to the logarithmic scaling in Fig.~\ref{fig:Agnostic_lin}, this is not immediately evident in the figure).
Our mode-addition procedure has reached convergence and no more additional modes need to be included.

\begin{figure}[h]
    \includegraphics[width = 0.9\columnwidth]{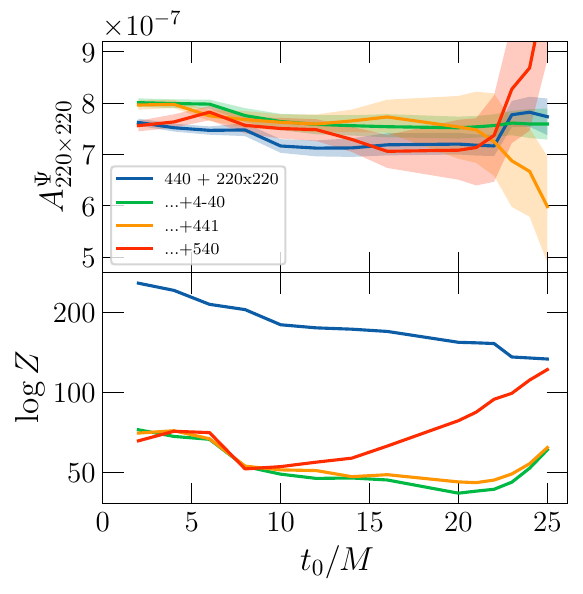}
    \caption{Amplitude of the $220\times 220$ mode (top) and Bayesian evidence $-\log Z$ as a function of the starting time of the fit, $t_0$, for the same configuration as in the previous figure. The bottom panel shows clearly that adding modes beyond the $440,\,4\text{-}40,\,220\times 220$ reduces the Bayesian evidence, and the amplitude of the quadratic mode extracted becomes less stable, with larger oscillations.}
    \label{fig:Agnostic_scd}
\end{figure}

The same method is then applied to the quadratic sector, this time quantifying the impact on the extraction of the $220\times220$ mode amplitude.
In this case, comparatively larger variations of the amplitudes are present, which are however compatible with the wide error bars (dictated by the numerical resolution), shown as shaded regions in Fig.~\ref{fig:Agnostic_scd}. 
Again, saturation of the evidence is observed for all times considered. Remarkably, this occurs when only the fundamental prograde and retrograde mode $440$ and $4\text{-}40$ are present, in addition to the quadratic mode $220\times 220$. Including any other mode only decreases the Bayesian evidence.

This procedure sets the ``default'' set of modes listed in Table~\ref{tab:modes} and used in the main text.
In order to correctly resolve the physical QNM content of the system in the quadratic sector, we must take into account 7 modes (corresponding to 14 free parameters).

Clearly, the time-variation discussed above is more relevant for modes in the quadratic sector.
For a pure QNM model to be valid, it is not sufficient for the $220$ and $220\times220$ amplitudes to be constant within the considered time window (within the estimated uncertainties): \textit{all the other modes} must also have approximately constant amplitudes~\cite{Baibhav:2023clw}.
This is why we often found it necessary to restrict the analysis to $t_0/M > 15$: at earlier times some of the modes exhibit large variations, which are incompatible with the numerical error.

\subsection{Error budget estimation}
\label{app:error_budget}

It is well-known that the starting time of the analysis has a significant influence on the QNM amplitude estimation, especially for the overtones~\cite{Dorband:2006gg,London:2014cma, London:2018gaq, Baibhav:2023clw}. 
Systematic uncertainties (including unaccounted-for QNMs, contributions from the prompt response or late-time tails, and numerical artifacts) can affect the inferred amplitude.
The impact of these uncertainties can be mitigated by imposing that, in the QNM-dominated regime, the amplitudes and phases appearing in Eq.~\eqref{eq:Kerr_model} should be independent of the starting time of the fit~\cite{London:2018gaq}.

We take into account this additional uncertainty by ``marginalizing'' over the starting time in the following way. For each run, we extract the amplitude posteriors at starting times $t_0/M = \{15, 17.5, 20, 22.5, 25\}$. 
Earlier times are excluded because we observe large variations in the parameters, indicating that a pure QNM description is not applicable yet.
The final estimate of the amplitude is then constructed by combining all of the posterior samples from each of the inferences, and extracting the mean and standard deviation of the combined posterior samples. This amounts to considering each of the inferences as part of the same random process. Such procedure results in a conservative  estimation of the error bars, since small variations due to the starting times or mode inclusion will enlarge the tails of the distribution.
We apply the above procedure, instead of a simple ``marginalization'' of the starting time as an additional parameter, because the template we consider models only a portion of the signal.
If the starting time were left free to vary as an additional parameter, the template would latch to the initial burst to increase the recovered power, and bias the mode extraction.
Tailored solutions to this issue might be considered by constructing tight prior intervals to marginalize the starting time~\cite{Carullo:2019flw, Finch:2022ynt}.
However, the results would be highly dependent on the template used (many-mode templates would latch more to the burst) and on the portion of the signal considered, thus not general and prone to systematics.
For this reason we prefer to apply the solution described above, which is simpler to implement.

\begin{figure}[ht!]
    \includegraphics[width = 0.9\columnwidth]{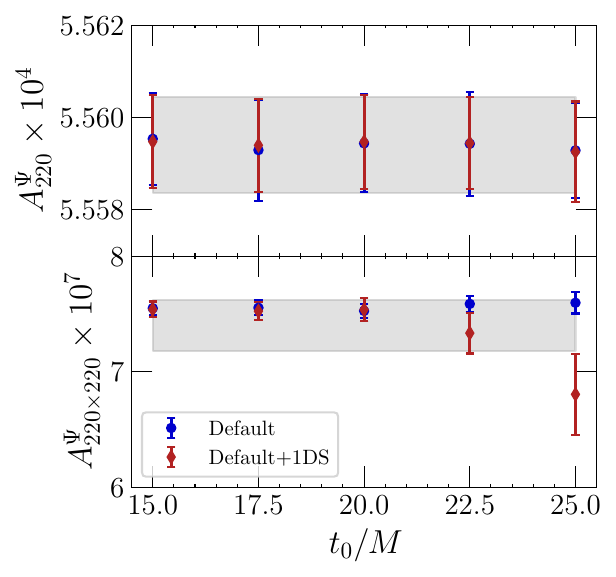}
    \caption{Extracted amplitude for the linear fundamental mode $(220)$ (top) and for the dominant quadratic mode $220\times 220$ (bottom), as a function of the starting time of the fit $t_0$, for a BH with spin $a/M = 0.7$. The red points show the extracted amplitudes, with their uncertainty estimates, for a template containing the default configuration; the blue points include one additional free damped sinusoid. The dashed gray regions are the extracted averaged amplitudes and error bars for each mode. 
    }
    \label{fig:Uncertainties}
\end{figure}

Capturing the full signal content for all the initial conditions considered is computationally prohibitive.
Besides, missing subdominant contributions with amplitudes close to the numerical error does not appreciably affect the extraction of the dominant QNM amplitudes in the large majority of cases.
However, for low spins, subleading counterrotating modes become more relevant, and they may affect the estimate of the QQNM amplitude.
We conservatively augment our error budget to account for this systematic uncertainty.
Whenever we use a template given by the superposition of a certain set of modes, we repeat the analysis using the same template plus an additional ``free'' mode (for which we do not specify the frequency). 
This ``free mode'' analysis yields a second estimate of the amplitude of each mode and of its uncertainty.
We conservatively (see discussion above) construct the final median estimate of our parameters as the median of the combined posterior samples of each of the inferences. 

In Fig.~\ref{fig:Uncertainties} we exemplify the construction of the two error budgets described above by focusing on the main quantities of interest in this study: the amplitude of the $220$ mode and of its quadratic counterpart (note the different y-axes).
The red markers represent the amplitude estimate (median and 90\% credible levels) for a certain ``default'' set of modes, at different time steps;
the blue markers represent the same amplitude estimate obtained by adding a free mode.
The starting time uncertainty and the free mode inclusion have negligible impact on the linear mode amplitude (with relative variations $\sim 10^{-4}$) , while their effect on the quadratic mode is of order a few percent -- i.e., large enough to be significant.

\subsection{Convergence of the numerical code\label{app:code_convergence}}

The simulations were run with a spin-dependent base grid resolution, because higher resolution is needed as the BH approaches extremality. For each run we used two different (but close-by) resolutions. The number of radial and angular points used in the runs, $\left(N_x, N_y\right)$, is listed in Table~\ref{tab:resolution}. 

\begin{table}[]
\caption{Resolution levels $(N_x, N_y)$ for different BH spins.}
\vspace{0.2cm}
{\begin{tabular}{ccc}
\hline \hline
$a/M$     & Low Resolution                    & High Resolution              \\
\hline
$0 \dots 0.5$     & $(184, 24)$ & $(186, 24)$   \\
$0.5 \dots 0.9$     & $(190, 24)$ & $(192, 24)$ \\ 
$0.9 \dots 1$     & $(196, 28)$ & $(198, 28)$ \\ 
\hline \hline
\end{tabular}
}
\label{tab:resolution}
\end{table}

\begin{figure}[t]
    \includegraphics[width = 0.9\columnwidth]{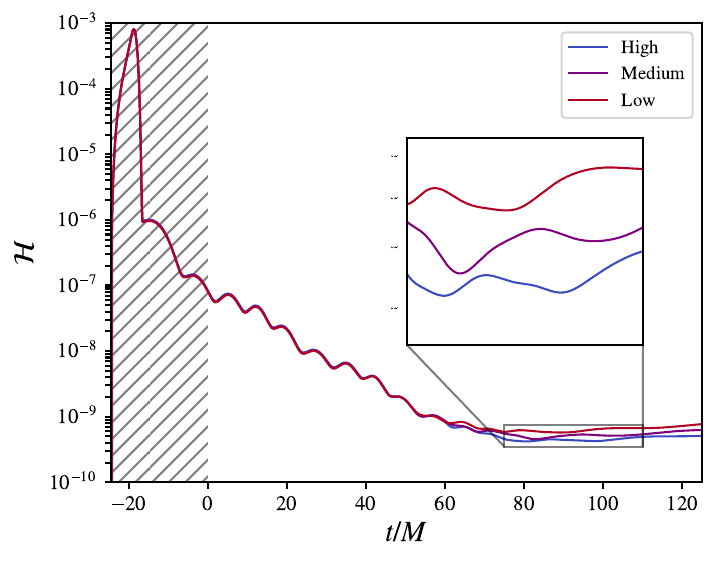}
    \caption{Independent residual stemming from the reality of the reconstructed metric components $\mathcal{H}$ (see Ref.~\cite{Ripley:2020xby}) as a function of time, for a BH spin $a/M = 0.7$, and for high $(N_x, N_y) = (192, 24)$, medium $(N_x, N_y) = (190, 24)$ and low $(N_x, N_y) = (188, 24)$ resolutions. The hatched region at $t < 0$ (where $t = 0$ is the peak of the second order perturbations) is not used for the Bayesian inference. At times $t > 0$ the residuals are small and converging to zero. The inset shows that the residuals are smaller for higher resolutions as expected, although the difference is small, because the resolutions are very close to each other.}
    \label{fig:Bianchi}
\end{figure}
\begin{figure*}[bh!]
    \centering
    \includegraphics[width = 0.9\textwidth]{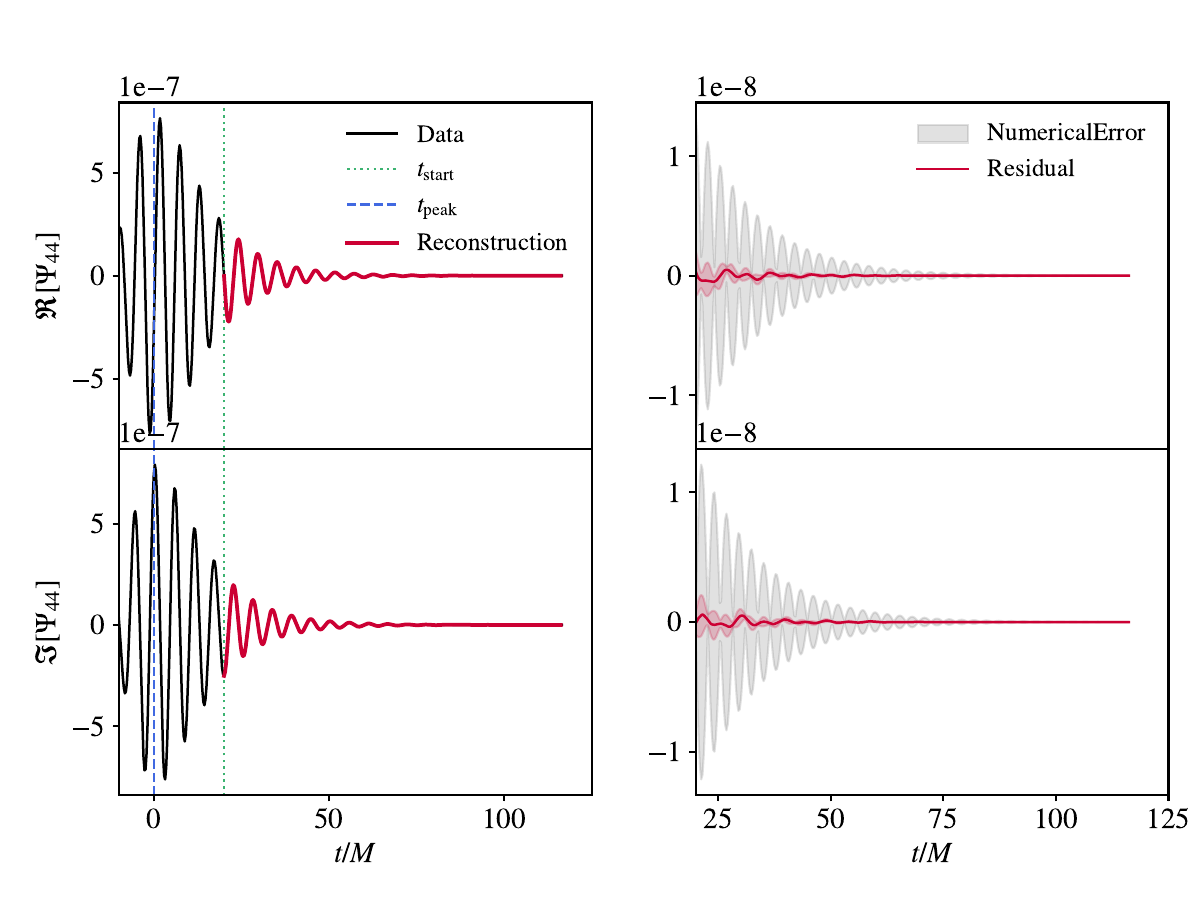}
    \caption{Left: The top (bottom) panels show the real (imaginary) part of the numerical data, in black, and the reconstructed waveform, in red. Amplitudes values reported in the text are referred to the time represented by the black dashed line, which is the peak of the second order perturbations. However, the template is only applied after the green dotted line, which is the starting time of our fit, where our QNM model is valid. This corresponds to the highest resolution among those considered in Fig.~\ref{fig:Bianchi}. Right: The top (bottom) panels show the residual between the real (imaginary) part of the reconstructed waveform and numerical data. The shaded red bands represent the $90\%$ confidence intervals in the reconstructed waveform. The gray shaded region is the envelope of the difference between two numerical resolutions. 
    }
    \label{fig:Reconstruction}
\end{figure*}
In order to test the numerical convergence of our code, we fix the BH spin to $a/M = 0.7$ and we assume initial conditions such that $r_l = 1.5r_h$, $r_u = 2.5 r_h$. We consider three runs with
$\left(N_x, N_y\right) = \left(188, 24\right),\left(190, 24\right),\left(192, 24\right)$. In Fig.~\ref{fig:Bianchi} we show an independent residual test for these three runs. The residual $\mathcal{H}$ is related to the components of the metric being real-valued functions (see Sec.~V.H of Ref.~\cite{Ripley:2020xby} for its definition). As usual, the time axis in the figure is such that $t = 0$ denotes the peak of the second-order perturbations. Constraint violations leave our computational domain at times $T_R < 0$ (i.e., in the region marked by hatched gray lines); at later times, the residuals converge rapidly to zero. Higher resolutions generally lead to smaller values of the residuals, as expected. Notice that the chosen resolutions are very close to each other and the convergence of the code is exponential, therefore the difference between resolutions is only visible in the inset. 

Even though the residuals are small and converging to zero at times $t \geq T_R$, some truncation error is inevitably present in the simulations. We estimate the truncation error by considering the source of numerical error in our inference to be the difference between the high and low resolutions defined in Table~\ref{tab:resolution}. The residual in the reconstructed waveform (see left panels of Fig.~\ref{fig:Reconstruction}) should be smaller than this numerical error. This is shown explicitly for one particular run in the right panels of Fig.~\ref{fig:Reconstruction}. The reconstructed residuals from the inference are shown in red (with error bars represented by shaded regions), while the difference between the two resolutions considered -- in this case, $(N_x, N_y) = (190, 24)$ and $(192, 24)$ -- is shown in gray. We observe that the residual error is well below the numerical noise.

\end{document}